\documentclass[aps,prd,10pt,a4paper,altaffilletter,amssymb,showpacs,nofootinbib,twocolumn,showkeys,a4paper,floatfix]{revtex4-1}
\usepackage{graphicx,epsfig}
\usepackage{subfig}
\usepackage{amsmath}

\begin{document}

\title{First cosmological constraints on the Superfluid Chaplygin gas model}
\author{Ruth Lazkoz$^{1}$, Ariadna Montiel$^{2}$ and Vincenzo Salzano$^{1}$ }

\affiliation{$^{}$ Dpto. de F\'{\i}sica Te\'orica, Universidad del Pa\'{\i}s Vasco, Apdo. 644, E-48080, Bilbao, Spain.\\
$^{2}$ Dpto. de F\'{\i}sica, Centro de Investigaci\'on y de Estudios Avanzados del I. P. N., Apdo. 14-740, D.F., Mexico.}

\begin{abstract}
In this work we set observational constraints of the Superfluid Chaplygin gas model, which gives a unified description of the dark sector of the Universe as a Bose-Einstein condensate (BEC) that behaves as dark energy (DE) while it is in the ground state and as dark matter (DM) when it is in the excited state. 
We first show and perform the various steps leading to a form of the equations suitable for the observational tests to be carried out. Then,
by using a Markov Chain Monte Carlo (MCMC) code, we constrain the model with a sample of cosmology-independent long gamma-ray bursts (LGRBs) calibrated using their Type I Fundamental Plane, as well as the Union2.1 set and observational Hubble parameter data. In this analysis, using our cosmological constraints, we sketch the effective equation of state parameter and deceleration parameter, and we also obtain  the redshift of the transition from  deceleration to acceleration: $z_t$.
\end{abstract}


\maketitle

\section{Introduction}

Understanding present cosmic acceleration is one of the major challenges in modern cosmology. This open problem has sparked an intense activity directed to investigate the possible candidates which fuel this acceleration. Typically, this accelerated expansion has been attributed to dark energy (DE) which makes up nearly 75$\%$ of the cosmic substratum. Another unknown component of the universe is the dark matter (DM), the missing mass necessary to held together galaxy clusters and also needed to explain the current large scale structure of the universe.

In the literature there are a considerable number of dark energy models; for a selection of DE models see \cite{Copeland}. Between those based on scalar field models we find are quintessence \cite{Wette88}, k-essence \cite{Armendariz2000}, phantom \cite{Cadwell}, and so forth. Another possibility to obtain accelerated expansion is provided by theories with large extra dimensions known as braneworlds \cite{Randall}. Furthermore, there is a stream that considers DM and DE as different manifestations of the same component and describes the dark sector as some kind of fluid whose physical properties depend on the scale: it behaves as DM at high densities and transforms into DE at lower ones. Most of these Unified Dark Matter models invoke the generalized Chaplygin gas (GCG), a perfect fluid characterized by a negative pressure which is inversely proportional to the energy density, as its paradigmatic example \cite{Kamenshchik:2001, Bilic:2002, Bento:2002}, but in this same vein we also find more phenomenological proposals \cite{bruni}.

Regarding unified dark matter models, recently in \cite{Popov} the author has proposed a superfluid Chaplygin gas (SCG) as a possible unification of both dark sector components: when in the state of a (background) Bose-Einstein condensate (BEC) it behaves like DE, while its excited states act as DM. The condensate possesses the equation of state of the Chaplygin gas, but the evolution of the Universe provided by this matter is different from that in the two-component model with the Chaplygin gas and cold dark matter (CDM) as well as from the GCG model used to unify DE and DM.

In \cite{Popov} it is showed that if $\nu$, the parameter that governs the DM equation of state, is quite large, the model will be consistent with cosmological observations, which implies that the pressure of normal component is small, and thus the superfluid behaves like CDM. From the superfluid standpoint it means that the second sound speed (name for the effective sound speed in the superfluid theory) is small, a characteristic that should be taken into account for any DM equation of state. On the other hand, if $k$, the parameter interrelating the Bose-Eistein condensate (DE) and the excited states (DM), can be estimated, it can be used to infer the redshift $z_t$ of the deceleration/acceleration  transition because it is directly related to the scale factor.

The good properties of the SCG model proposed in \cite{Popov}, and the subsequent analysis from a state-finder point of view developed in \cite{Popov11}, justifies to carry out a observational confrontation using the most recent compilation of supernovae Ia (SNe Ia) Union2.1 \cite{Union21} to explore as first approach its reach and faithfulness. Besides SNe Ia, we use the updated sample of Hubble parameter data \cite{Jimenez12} and the distance modulus of long Gamma-Ray Bursts (LGRBs) \cite{Yonetoku12}.

The work is structured as follows. In Sec. II, we introduce the main aspects of the SCG model and discuss how the equations are to be rearranged for the cosmological constraints analysis. Next, in Sec. III, we describe the observational data samples chosen for our analysis: SNe Ia, Hubble parameter estimates and GRBs. In Sec. IV we discuss the constraints and results obtained, and we point out our concluding remarks.

\section{The SCG model with FRW spacetime}

The SCG model is based on the action
\begin{equation}
 S=\int \left( -\frac{R}{16\pi G} + \mathcal{L} \right) \sqrt{-g} d^4x,
\label{Eq:action}
\end{equation}
where the Lagrangian $\mathcal{L}$ is associated to the generalized hydrodynamic pressure function depending on a single variable if it is considered a pure condensate and on three variables if the gas excitations are included.

In \cite{Popov} it is argued that the two-fluid hydrodynamics constitutes an efficient scheme to describe excited states basically because it does not depend on details of microscopic structure of the quantum liquid and then capitalizes on effective macroscopic quantities.

The superfluid and normal component are two independent flows, and due that it is necessary to increase the number of independent variables in the generalized pressure from one to three \cite{Khalat}. The latter leads to three scalar invariants that can be constructed from the pair of independent vectors: the superfluid $\mu_{\alpha}$ and thermal $\theta_{\alpha}$ momentum covectors allowing to write the general variation of the generalized pressure in a fixed background as $\delta P=\delta \mathcal{L}=n^{\alpha}\delta \mu_{\alpha}+s^{\alpha}\delta \theta_{\alpha}$ with $n^{\alpha}$ and $s^{\alpha}$ the particle number and entropy currents respectively.

The energy-momentum tensor, in terms of the 4-velocities $U^{\alpha}$ and $V^{\alpha}$, is given by 
\begin{equation}
 T_{\alpha \beta}= \mu n_c V_{\alpha} V_{\beta} + W_n U_{\alpha} U_{\beta} -P g_{\alpha \beta}. 
\label{Eq:EMTensor}
\end{equation}

Besides, the particle number current is
\begin{equation}
n^{\alpha}=n_c V^{\alpha} + n_n U^{\alpha}.
\label{Eq:consnum}
\end{equation}

The SCG model considers the cosmic substratum as matter in the BEC state with energy-momentum tensor and particle number current given by Eqs. (\ref{Eq:EMTensor}) and (\ref{Eq:consnum}) respectively, and also assumes the following self-dependent condensate ansatz
\begin{equation}
P(\mu,\beta,\gamma)= p_c(\mu) + p_n (\mu,\beta,\gamma) \; ;
\end{equation}
besides that the superfluid background obeys the equation of state
\begin{equation}
n_c=\sqrt{\frac{\lambda}{A}}\sqrt{\rho^2_c-A}, \qquad p_c=-\frac{A}{\rho_c},
\end{equation}
with the excited state described by the relations
\begin{equation}
\mu \gamma n_n=(1-c^2_s)W_n, \qquad p_n=c^2_s W_n/(1+\nu),
\label{Eq:Pdm}
\end{equation}
where $\mu$ is a chemical potential, $\gamma$ a scalar associated with the relative motion of the components and $c_s$ the adiabatic speed of sound.

For a homogeneous and isotropic spatially flat universe, the superfluid and normal velocities are equal and thus $\gamma=1$. In this case, the Einstein equations reduce to the Friedmann ones:
\begin{equation}
\label{EinsteinEqs}
3\frac{\dot a^2}{a^2} = 8\pi G \rho_{\textrm{tot}},
\qquad
\frac{\ddot a}{a} = -\frac{4 \pi G}{3}  (3p_{\textrm{tot}}+\rho_{\textrm{tot}}),
\end{equation}
where $\rho_{\textrm{tot}}$ consists of the condensate density, $\rho_{\textrm{c}}$, and the normal one, $\rho_{\textrm{n}}=W_{\textrm{n}}-p_{\textrm{n}}$, that are interpretable as DE and DM densities respectively, and $p_{\textrm{tot}}=p_{\textrm{c}}+p_{\textrm{n}}$.

As usual, the total energy density satisfies a conservation law:
\begin{equation}
\label{eq:conservation}
\dot\rho_{\textrm{tot}} + 3\frac{\dot a}{a}(p_{\textrm{tot}}+\rho_\textrm{tot})=0,
\end{equation}
which in this case contains implicitly the interaction between DE and DM, expressed by the particle number conservation,
\begin{equation}
\label{eq:numberconservation}
\dot n_{\textrm{tot}} + 3\frac{\dot a}{a}n_{\textrm{tot}}=0.
\end{equation}

In \cite{Popov}, from the Friedmann and conservation law equations, the following two dimensionless expressions are obtained:
\begin{eqnarray}
 &&%
 3(1+\nu)\frac{\dot a^2}{a^2} = \frac{1}{\rho}+
\frac{k}{a^3}\left(\frac{\nu\rho}{\sqrt{\rho^2-1}}+\frac{\sqrt{\rho^2-1}}{\rho}\right),
\label{Eq2a}\\ &&%
3\frac{\dot a}{a}\left(1+\nu-
            \frac{k}{a^3}\frac{1}{\sqrt{\rho^2-1}}\right) + \nonumber \\ &&
            \frac{\dot\rho}{\rho}
            \left(1- \frac{k}{a^3}\left(\frac{1}{\sqrt{\rho^2-1}}-
            \frac{\nu\rho^2}{(\rho^2-1)^{3/2}}\right) \right)=0,
\label{Eq2b}
\end{eqnarray}
where $\rho=\rho_{\rm c}/\sqrt{A}$, and $k=n_0/\sqrt{\lambda}$, the parameter $k$ giving an initial normalized total particle number density.

Clearly, for any given value of the parameter $\nu$ we would have a different solution for the fluid dynamics; we note that in the limit we are interested, $\nu \to\infty$, the quasiparticle pressure, Eq. (\ref{Eq:Pdm}), is zero, thus behaving as dust-like matter ($p_n=0$), and Eqs. (\ref{Eq2a},\ref{Eq2b}) can be solved analytically. In this case, Eqs.~(\ref{EinsteinEqs})~-~(\ref{Eq2b}) yield to
\begin{equation}
\rho_{\rm c} = \sqrt{\frac{k^2}{(a^3+\kappa_0)^2}+1},
\label{Eq:rhoc}
\end{equation}
while DM is governed by
\begin{equation}
\rho_{\textrm{n}} = \frac{\kappa_0}{a^3}
\sqrt{\frac{k^2}{(a^3+\kappa_0)^2}+1}.
\label{Eq:rhon}
\end{equation}
The parameter $\kappa_0$ is clearly associated to the ratio of normal and condensate density evaluated at present.

From Eqs. (\ref{Eq:rhoc}) and (\ref{Eq:rhon}) it can be noted that at the beginning stage (i.e. for small $a$) the total energy density is approximated by $\rho_{\textrm{tot}}\propto a^{-3}$ corresponding to a universe dominated by dust-like matter. In \cite{Popov} it is pointed that although the same behavior is a feature of  Chaplygin gas \cite{Kamenshchik:2001} and that even though in this model the condensate has the same equation  of state, such  dependence is due to the normal component.

With Eqs. (\ref{Eq:rhoc}) and (\ref{Eq:rhon}), the first Friedmann equation turns out to be
\begin{equation}
3H^2=8\pi G \left[1+\frac{\kappa_0}{a^3} \right] \sqrt{\frac{k^2}{(a^3+\kappa_0)^2} + 1},
\end{equation}
and after some simple algebra, we can rewrite it as:
\begin{equation}
\frac{H^2}{H_0^2}=(1+\kappa_0)\frac{\Omega_{c}}{a^3}\sqrt{\frac{k^2 + (a^3+\kappa_0)^2}{k^2+(1+\kappa_0)^2}}.
\label{eq:fri}
\end{equation}
where, as usual, $\Omega_{c,0} = {8 \pi G}/({3 H_{0}^2} \rho_{c,0})$. Henceforth, $\Omega_{c} \equiv \Omega_{c,0}$ just for simplicity. It is straightforward to derive that, demanding $H(z=0)=H_0$, there is a relation between $\Omega_{c}$ and $\kappa_{0}$, it being: $\Omega_{c}(1+\kappa_0)=1$. Using this condition, we can reduce the dimensionality of our problem, thus having only two free parameters, $\Omega_{c}$ and $k$:
\begin{equation}
 \frac{H^2}{H_0^2}=\frac{1}{a^3} \sqrt{\frac{k^2 \Omega^2_{c}+\left[1+ (a^3-1) \Omega_{c} \right]^2}{1+k^2 \Omega^2_{c}}}.
\label{Eq:F}
\end{equation}

On the other hand, it is known that the equation of state parameter is connected directly to the evolution of the energy density, and thus to the expansion of the universe. In the case of the SCG model, the knowledge of $k$ and $\kappa_0$ suffice to determine $w$ as follows: 
\begin{equation}
w=\frac{p_{tot}}{\rho_{tot}}= -\frac{a^3(a^3+\kappa_0)}{k^2+(a^3+\kappa_0)^2}.
\end{equation}
Note that $w$ goes to $-1$ when $a \rightarrow \infty$ and $w$ goes to $0$ when $a\rightarrow 0$.

Finally, we use the deceleration parameter for the SCG model given by
\begin{equation}
 q=\frac{1}{2} \left[1-\frac{3a^3(a^3+\kappa_0)}{k^2+(a^3+\kappa_0)^2} \right],
\label{Eq:q}
\end{equation}
to obtain the transition redshift $z_t$ and thus know the prediction of the SCG model for the cosmic acceleration start.


\section{Observational tests}

In the following, we use the observational data from Union2.1 supernovae data set, the update of Union2 reported in \cite{Union21}, the observational Hubble parameter data given in \cite{Jimenez12}, the distance modulus of long Gamma-ray Bursts reported in \cite{Yonetoku12} to constrain the two free parameters of the model ($\Omega_c$, $k$). In order to do that, we use a Markov Chain Monte Carlo (MCMC) code to maximize the likelihood function $\mathcal{L}(\theta_i) \propto \exp [-\chi^2(\theta_i)/2]$ where $\theta_i$ is the set of model parameters and the expression for $\chi^2(\theta_i)$ depends on the dataset used. The MCMC methods (completely described in \cite{Berg,MacKay,Neal} and references therein) are well-established techniques for constraining parameters from observational data. To test their convergence, here we follow the method developed and fully described in \cite{Dunkley05}.

\subsection{Supernovae}

SNe Ia are considered standard candles, so they can be used to measure directly the expansion rate of the universe. Recently, the Supernova Cosmology Project (SCP) collaboration released the updated Union2.1 compilation which consists of 580 SNe Ia \cite{Union21}. The Union2.1 compilation is the largest published and spectroscopically confirmed SNe Ia sample to date. Constraints from the SNe Ia data can be obtained by fitting the distance moduli $\mu(z)$. A distance modulus can be calculated as
\begin{eqnarray}
\mu(z_j)&=& 5 \log_{10} [D_L(z_j, \theta_i) ] + 25 \nonumber \\
        &=& 5 \log_{10} [d_L(z_j, \theta_i) ] + \mu_0,
\label{Eq:mu}
\end{eqnarray}
where $\mu_0=42.38-5\log_{10} h$ and $h$ is the Hubble constant $H_0$ in units of 100 km s$^{-1}$Mpc$^{-1}$, whereas $d_L(z_j,\theta_i)$ is the Hubble free luminosity distance defined as
\begin{equation}
d_L(z,\theta_i)\equiv H_0 D_L(z,\theta_i)= (1+z)\int^z_0  \frac{dz'}{E(z',\theta_i)},
\end{equation}
in which $E(z,\theta_i)=H(z,\theta_i)/H_0$ and $\theta_i$ denotes the vector model parameters.

The $\chi^2$ function for the SNe Ia data is
\begin{equation}
\chi^2_\mu (\mu_0, \theta_i)= \sum^{580}_{j=1} \frac{(\mu(z_j; \mu_0, \theta_i)-\mu_{obs}(z_j))^2}{\sigma^2_{\mu}(z_j)},
\end{equation}
where the $\sigma^2_{\mu}$ corresponds to the error on distance modulus for each supernova. The parameter $\mu_0$ in the Eq. (\ref{Eq:mu}) is a nuisance parameter since it encodes the Hubble parameter and the absolute magnitude $M$ and because of that it has to be marginalized over. Here, we follow the alternative method suggested in \cite{Pietro03,Nesseris05} which consists  in maximizing the likelihood by minimizing $\chi^2$ with respect to $ \mu_0$. Then one can rewrite the $\chi^2$ as
\begin{equation}
\chi^2_{SN} (\theta)= c_1 - 2c_2 \mu_0 + c_3 \mu^2_0,
\end{equation}
where
\begin{equation}
c_1=\sum^{580}_{j=1} \frac{(\mu(z_j; \mu_0=0,\theta_i)-\mu_{obs}(z_j))^2}{\sigma^2_{\mu}(z_j)},
\end{equation}

\begin{equation}
c_2=\sum^{580}_{j=1} \frac{(\mu(z_j; \mu_0=0,\theta_i)-\mu_{obs}(z_j))}{\sigma^2_{\mu}(z_j)},
\end{equation}

\begin{equation}
c_3=\sum^{580}_{j=1} \frac{1}{\sigma^2_{\mu}(z_j)}.
\end{equation}

The minimization over $\mu_0$ gives $\mu_0=c_2/c_3$. So the $\chi^2$ function takes the form
\begin{equation}
\tilde{\chi}_{SN} (\theta_i)= c_1 - \frac{c^2_2}{c_3}.
\end{equation}

Since $\tilde{\chi}^2_{SN}=\chi^2_{SN}(\mu_0=0,\theta_i)$ (up to a constant), we can instead minimize $\tilde{\chi}^2_{SN}$ which is independent of $\mu_0$.

\subsection{Hubble parameter observations}

We use the compilation of Hubble parameter measurements estimated with the differential evolution of passively evolving early-type galaxies as “cosmic chronometers”, in the redshift range $0<z<1.75$ recently updated in \cite{Jimenez12} but first reported in \cite{Jimenez02}.

The root idea supporting this approach is the measurement of the differential age evolution of these chronometers as a function of redshift, which provides a direct estimate of the Hubble parameter $H(z) = -1/(1 + z)dz/dt \simeq -1/(1 + z)\Delta z/\Delta t$. The main strength of this approach is the confidence on the measurement of a differential quantity, $\Delta z/ \Delta t$, which provides many advantages in minimizing many common issues and systematic effects. In addition, compared with other techniques, this approach provides a direct measurement of the Hubble parameter, and not of its integral, in contrast to SNe Ia or angular/angle-averaged BAO.

Observed values of $H(z)$ can be used to estimate the free parameters of the model and also the best-value for $H_0$ by minimizing the quantity
\begin{equation}
\chi^2_H (H_0, \theta_i)= \sum^{18}_{j=1} \frac{(H(z_j; \theta_i)-H_{obs}(z_j))^2}{\sigma^2_H(z_j)},
\end{equation}
where $\sigma^2_H$ are the measurement variances. In addition, we fix $H_0$ at the value given in \cite{Riess}, $H_{0} = 73.8\pm2.4$. The vector of model parameters, $\theta_i$, will be $\theta_i=(\Omega_c, k)$.
\begin{table*}
\begin{ruledtabular}
\begin{tabular}{ccccccc}
                       & $\Omega_c$               & $k$                        & $\chi^2_{red}$& $k_0$ & $z_t$  \\ \hline
SNe Ia                  & $0.774^{+0.085}_{-0.047}$& $0.278^{+0.116}_{-0.161}$  & 0.975& $0.291^{+0.079}_{-0.142}$& $0.654^{+0.175}_{-0.200}$\\
Hubble                 & $0.845^{+0.094}_{-0.076}$& $0.282^{+0.065}_{-0.131}$  & 0.913& $0.183^{+0.107}_{-0.132}$& $0.721^{+0.130}_{-0.230}$\\
GRBs                   & $0.869^{+0.071}_{-0.051}$& $0.174^{+0.077}_{-0.099}$  & 0.846& $0.151^{+0.067}_{-0.094}$& $0.977^{+0.238}_{-0.277}$  \\
SNe Ia + Hubble + GRBs  & $0.771^{+0.046}_{-0.025}$& $0.173^{+0.109}_{-0.108}$  & 0.976& $0.297^{+0.043}_{-0.078}$& $0.772^{+0.160}_{-0.135}$  \\
\end{tabular}
\caption{Summary of the best estimates of model parameters as well as for the transition redshift (deceleration/acceleration) $z_t$.  The errors are at 68.3$\%$ confidence level. See Figure \ref{Fig:Contours}. }
\label{T:1}
\end{ruledtabular}
\end{table*}

\subsection{LGRBs}

The luminosity of GRBs appears to be correlated with their temporal and spectral properties. Although these correlations are not yet fully understood from first principles, their existence has naturally suggested the use of GRBs as distance indicators offering a possible route to probe the expansion history of the Universe. Unlike SNe Ia, GRBs can be detected at redshifts beyond $1.7$; however, the problem is that GRBs appear not to be standard candles and to extract cosmological information it is necessary calibrate them for each cosmological model tested.

There have been several efforts to calibrate the correlations between the luminosity and spectral properties of GRBs in a cosmology-independent way and among the proposals to use SNe Ia measurements to calibrate them externally are \cite{Kodama08,Liang08,Wei09}. Indeed, recently it has been made an update of the sample of GRBs now consisting of 109 objects \cite{Wei10} and new calibrations have been developed in order to use GRBs in cosmological tests \cite{Wei10,Montiel}. Nevertheless the cosmological constraints from GRBs are sensitive to how these are calibrated and if one calibrate them using Type Ia supernovae tighter constraints are obtained than calibrating them internally. In \cite{Wang08} GRBs have been calibrated internally, without using any external data sets, and although this method has the advantage to generate a sample that can be directly combined with other data to derive cosmological constraints, it has also the drawback that is not completely cosmology independent because it requires a cosmological model as input.

More recently, in the direction of the first method to calibrate GRBs, in \cite{Yonetoku12} it has been estimated the distance modulus to long gamma-ray bursts (LGRBs) using the Type I Fundamental Plane, a correlation between the spectral peak energy $E_p$, the peak luminosity $L_p$, and the luminosity time $T_L\equiv E_{iso}/L_p$, where $E_{iso}$ is isotropic energy. Basically the calibration was done in this way: first, the Type I Fundamental Plane of LGRBs was calibrated using 8 LGRBs with redshift $z<1.4$ and SNe Ia (Union2) in the same redshift range by a local regression method, to avoid any assumption on a cosmological model; then this calibrated Type I Fundamental Plane was used to measure the distance modulus to 9 high-redshift LGRBs (see \cite{Yonetoku12} for calibration's details). To derive constraints on cosmological parameters of the model concerned we use these 9 LGRBs calibrated reported in Table~2 of \cite{Yonetoku12}.

The $\chi^2$ function for the GRBs data is defined by
\begin{equation}
\chi^2(\theta_i )= \sum_{j=1}^{9} \left[\frac{\mu(z_j, \theta_i) - \mu_{\rm obs}(z_j)}{\sigma_{\mu_j}}\right]^{2}
\end{equation}
where $\mu(z_j)= 5 \log_{10} [d_L(z_j, \theta_i) ] + 25$. Notice that we have used the standard expression of $\chi^2$ given through the observed distance moduli just to be consistent with the way in which the calibration was done. We have fixed $H_0$ as $73.8\pm2.4$ from \cite{Riess}.

\section{Results and concluding remarks}

\begin{figure}
\includegraphics[width=0.45\textwidth]{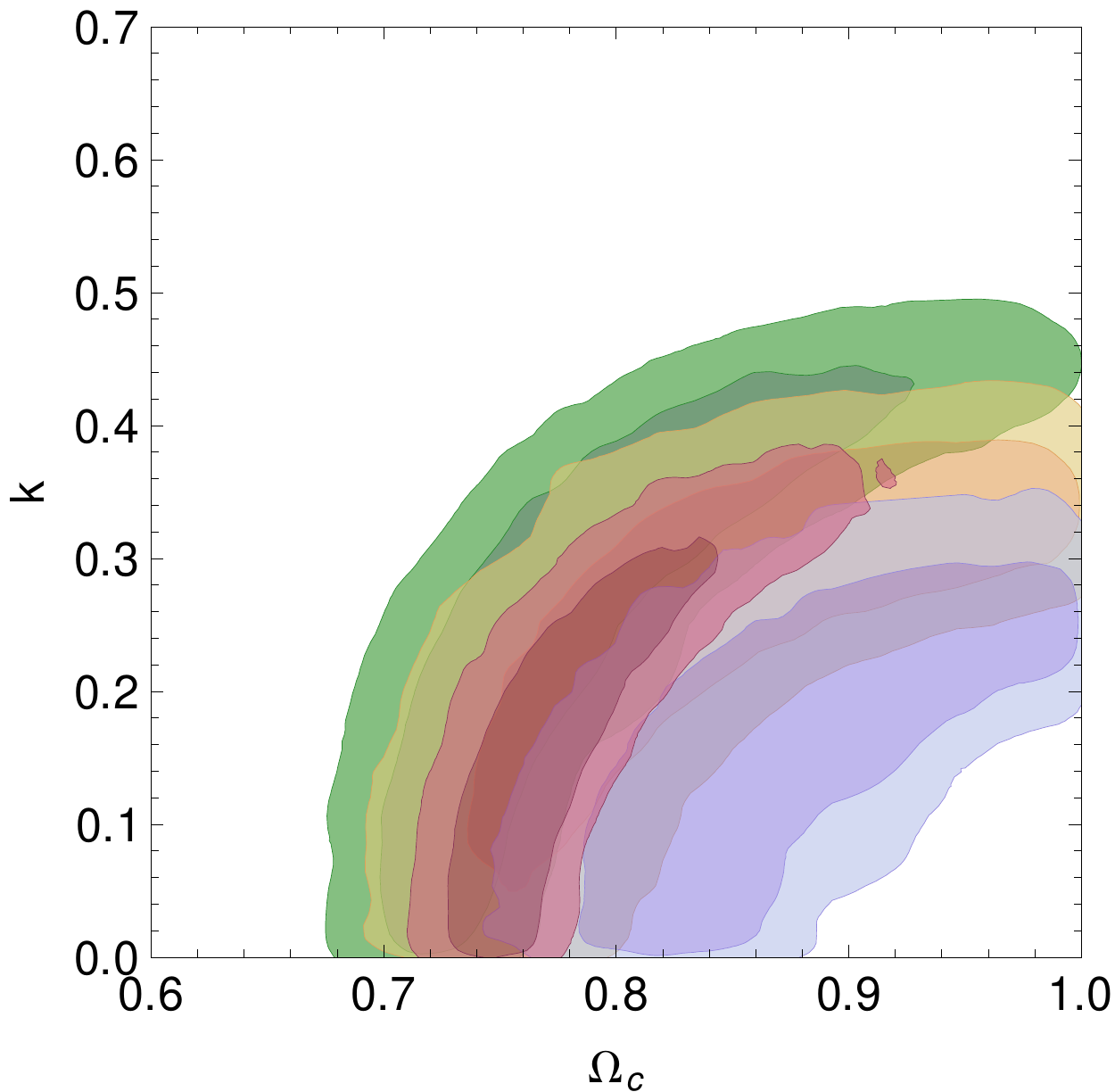}
\caption{Confidence regions in the ($\Omega_c$, $k$) plane for the SCG model. The contours correspond to 1$\sigma$-2$\sigma$ confidence regions using: SNe Ia, the green ones; Hubble parameter data, the yellow ones; LGRBs, the violet ones; the combination of all observational data, the red ones.}
  \label{Fig:Contours}
\end{figure}

\begin{figure}
\includegraphics[width=0.45\textwidth]{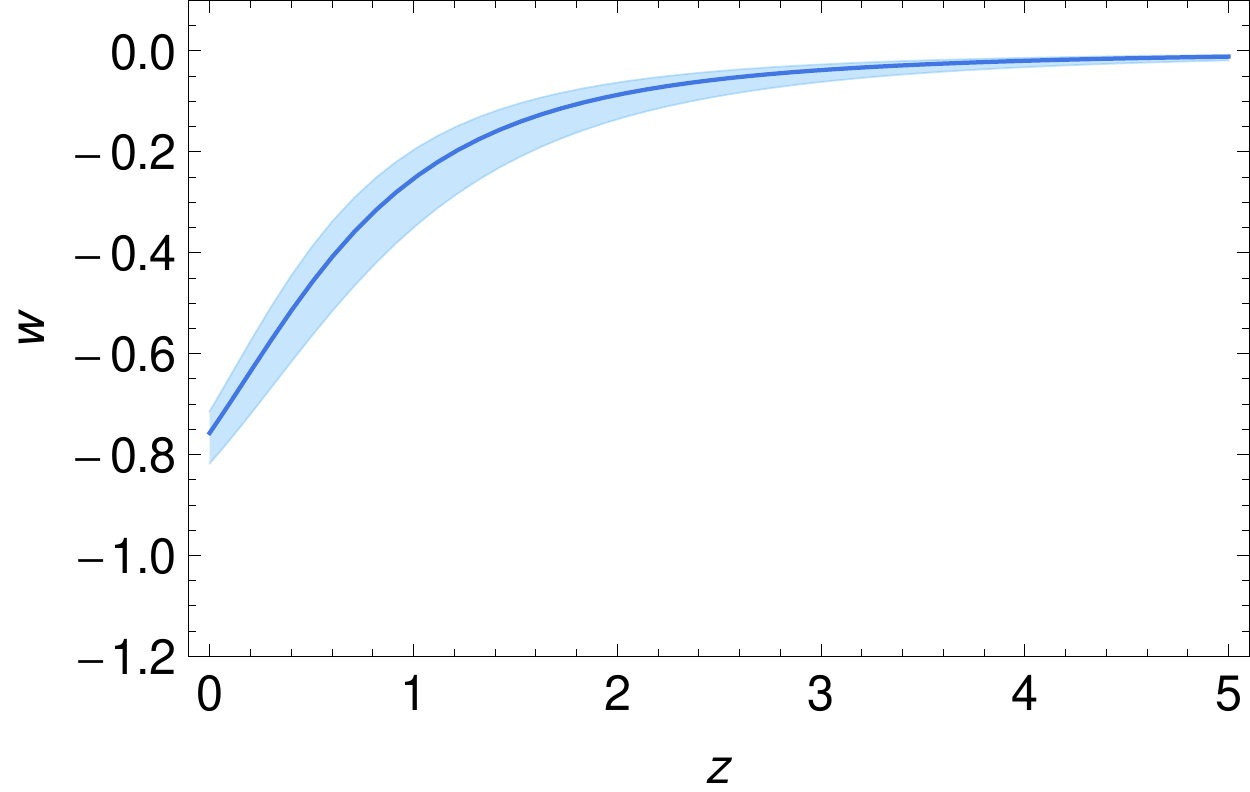}
\caption{Effective equation of state parameter $w(z)$ using the best fit for $k$ and $k_0$. The central line represents the best fit and the shaded contour represents the 1$\sigma$ confidence level around the best fit.}
  \label{Fig:w}
\end{figure}

\begin{figure}
\includegraphics[width=0.45\textwidth]{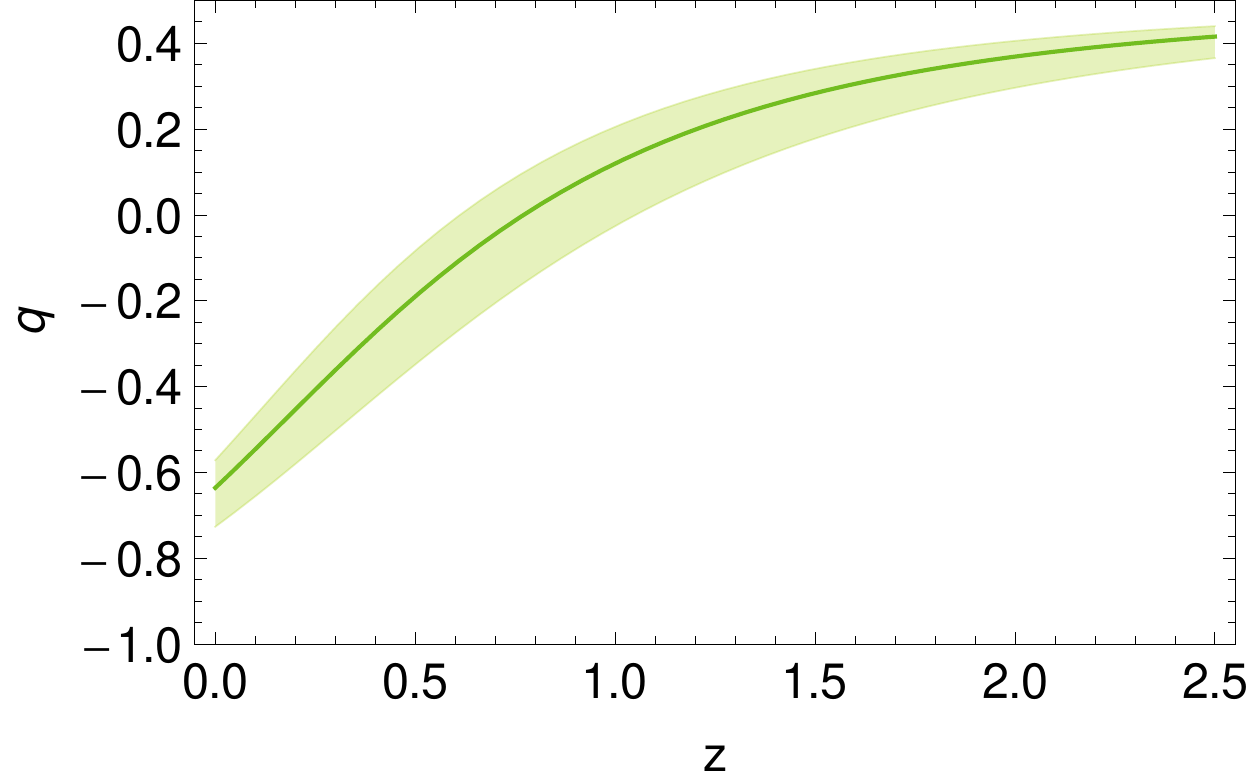}
\caption{Deceleration parameter $q(z)$ evolving with redshift using the best fit for $k$ and $k_0$. The central line represents the best fit and the shaded contour represents the 1$\sigma$ confidence level. }
  \label{Fig:q}
\end{figure}

As we have pointed above, to obtain the best fit parameter values of the SCG model, we applied the maximum likelihood method. The parameter space of the model is ($\Omega_c$, $k$) where $\Omega_c$ represents the fractional energy density for DE, and the value of $k$ is directly related to the scale factor corresponding to the transition from the deceleration to the acceleration epoch. Indeed, in \cite{Popov11} it has been found the restriction $k<0.652$ for the dust-like normal component, in order to be in agreement with the fact that the transition should occur at $0.3<z_t<1$ \cite{Riess04}. So, from our analysis we would like to know the prediction for the value of $k$ as well as of $\Omega_c$.

The best fit parameter values with $1\sigma$ error and the corresponding values of $\chi^2$ are summarized in the first three columns of Table~I and in Fig.~1 the $68\%$ and $95\%$ confidence level contours are displayed. As we can see from Fig.~1 and from the values for the $\chi^2_{red}$, the best-fits occur by using SNe Ia and the combination of all observational dataset. In Fig.~2 we present the evolution of the effective equation of state parameter $w(z)$ using our best fit. In this case and for current time, it turns out to be $w=-0.758^{+0.061}_{-0.043}$.

On the other hand, to obtain the transition redshift, it is necessary to know the value for $\kappa_0$, see Eq. (\ref{Eq:q}). To obtain this quantity, we use the normalization condition, from which using the respective best fit value for $\Omega_c$ from each observational dataset, we get the corresponding value for $\kappa_0$. The last two columns of Table~I shows our results for $\kappa_0$ as well as for $z_t$, which is a result of solving  the equation $q(z)=0$. Furthermore, in Fig.~3, the evolution of $q(z)$ with $z$ can be seen using once again our best fit. 

In our results, the best-fit parameters do not have symmetric errors, thus, we cannot use the standard error propagation formula and we have to perform a modification in order to account for these non-gaussianities. Here we follow the method developed in \cite{Lazkoz07} which we explain in the Appendix.

Our results can be summarized as follows:
\begin{itemize}
\item First, from the constraints using SNe Ia and the combination of all three observational datasets, we obtained an abundance of DE, given through $\Omega_c$ in the SCG model, in agreement with the prediction of the $\Lambda$CDM model from WMAP7 data alone, $\Omega_{\Lambda}=0.73\pm0.03$ \cite{wmap7}.

\item Second, when we use the Hubble parameter observations and LGRBs, we get restrictions slightly less tighter respect to the ones obtained using only SNe Ia; however, the amount predicted for $\Omega_c$ is not so far from the abundance expected. In the case of LGRBs, there probably still exist considerable systematic uncertainties which are reflected in the value of the $\chi^2_{red}$ and also in the respective confidence region.

\item Third, if we accept the fact that $k$ should be less than $0.652$ to guarantee a compatible value with $z_t$, as has been suggested in \cite{Popov11}, we can say that ours results are in total agreement with this restriction. In the $\Lambda$CDM model, the expansion of the universe switched from deceleration to acceleration at $z_t=0.752 \pm 0.041$ \cite{Union21} and in the SCG model, according to our results obtained from the joint of all three observational datasets, the universe exhibits the transition from deceleration to acceleration at $z_t=0.772^{+0.160}_{-0.135}$. However, the value for $z_t$ as a result of GRBs alone, is higher than the results from SNe Ia, Hubble parameter and the combination of all of them, but still agrees in $1\sigma$ confidence.

\item Finally, taking together SNe Ia, Hubble parameter and LGRBs, we obtained as best-fit model parameters ($\Omega_c=0.771^{+0.046}_{-0.025}$, $k=0.173^{+0.109}_{-0.108}$). Looking at the results shown in Table I, we will see that in this best-fit, the SNe Ia result dominates the constraint on $\Omega_c$, whereas the constraint on $k$ is visibly dominated by the GRBs result. This can be understood if we take in mind that $\Omega_c$ is a parameter evaluated at the present time, $a=1$, and that $k=n_0/\sqrt{\lambda}$ is evaluated at $a=0$, thus GRBs reasonably being a better tool to constrain $k$ than SNe Ia, whereas SNe Ia are better than GRBs to constrain $\Omega_c$. So, in conclusion, this result is affected by the redshifts at which the observational data set operate. On the other hand, comparing our best-fits with those obtained with SNe Ia, Hubble parameter and LGRBs by themselves, we see that the errors of model parameters are shrunk when LGRBs is used as a complementary cosmic constraint. Indeed, although the number of LGRBs is low, in \cite{Yonetoku12} it is pointed out that this sample yields to much stronger constraints due to the control of systematic errors which have even been removed thus enabling consistence with other probes, e.g. Cosmic Microwave Background (CMB) and Baryon Acoustic Oscillation (BAO). From our results, we can also  say that LGRB data are a valuable tool in probing models where dark matter and dark energy are unified if they are considered together with other observational datasets.

\end{itemize}

\appendix
\section{Error propagation}

The constraints on the parameters are given in the form ${\theta_i}^{+\delta\theta_{i,u}}_{-\delta\theta_{i,d}}$, where $\delta\theta_{i,u}$ and $\delta\theta_{i,d}$ are positive quantities.

The estimated error in a quantity depending on them, $f(\mathbf{\theta})$, will be given by an upper limit
\begin{equation}
\Delta f_u= \sqrt{\sum_{i=1}^n\left({\rm max}\left(\Delta f_{iu},-\Delta f_{il}\right)\right)^2},
\end{equation}
and a lower one
\begin{equation}
\Delta f_l= \sqrt{\sum_{i=1}^n\left({\rm min}\left(\Delta f_{iu},-\Delta f_{il}\right)\right)^2},
\end{equation}
where
\begin{equation}
 \Delta f_{iu}=f(\dots\theta_{(i-1)},\theta_{i}+\Delta\theta_{iu},
\theta_{(i+1)},\dots)-f(\mathbf{\theta})
\end{equation}
\begin{equation}
 \Delta f_{il}=f(\dots\theta_{(i-1)},\theta_{i}-\Delta\theta_{il},
\theta_{(i+1)},\dots)-f(\mathbf{\theta}).
\end{equation}

This error estimation is based on finite differences. If the errors are enough small, it can be refined: $\Delta\theta_{i,u}=\delta\theta_{i,u}$ and $\Delta\theta_{i,l}=\delta\theta_{i,l}$. In that case one can write
\begin{equation}
\Delta f_u\simeq\delta f_u= \sqrt{\sum_{i=1}^n\left({\rm max}\left(\frac{\partial f}{\partial \theta_i}\delta\theta_{iu},-\frac{\partial f}{\partial \theta_i}\delta\theta_{il}
\right)\right)^2}
\end{equation}
and
\begin{equation}
\Delta f_l\simeq\delta f_l= \sqrt{\sum_{i=1}^n\left({\rm min}\left(\frac{\partial f}{\partial \theta_i}\delta\theta_{iu},-\frac{\partial f}{\partial \theta_i}\delta\theta_{il}
\right)\right)^2}.
\end{equation}

In Gaussian situations, where $\Delta\theta_{i,u}=\Delta\theta_{i,l}=\Delta\theta_i$, one gets the standard error propagation formula and $\Delta f_u=\Delta f_l$.

\begin{acknowledgments}
RL and VS supported by the Spanish Ministry of Economy and Competitiveness through research projects FIS2010-15492 and Consolider EPI CSD2010-00064, and also by the Basque Government through research project   GIU06/37 respectively,
and by the University of the Basque Country UPV/EHU under program UFI 11/55 and special action ETORKOSMO. AM would like to thank the colleagues of UPV/EHU for kind hospitality as well as financial support by CONACyT (Mexico) through a \textit{Ph.D Beca Mixta Project}.
\end{acknowledgments}

\end{document}